\documentclass[a4paper,11pt]{article}
\usepackage{jheppub}
    \usepackage{etoolbox}
    \makeatletter
    \patchcmd{\maketitle}{\@fpheader}{}{}{}
    \makeatother
        
\usepackage[usenames,dvipsnames,table]{xcolor}
\usepackage[T1]{fontenc} 
\usepackage[utf8]{inputenc}
\usepackage{tikz, pgf}
\usepackage{amsmath, mathtools}
\usetikzlibrary{shapes.misc}
\usetikzlibrary{shapes}
\usetikzlibrary{fit}
\usetikzlibrary{arrows}
\usepackage{array,tikz-cd}
\usepackage{subfigure}
\usepackage{braket}
\usepackage{environ}
\usepackage{float}
\usepackage[shortlabels]{enumitem}

\NewEnviron{eqn}{
\begin{align}
\begin{split}
  \BODY
\end{split}
\end{align}
}

\definecolor{rust}{rgb}{0.8,0.2,0.2}

\newcommand{\be}{\begin{equation}}
\newcommand{\ee}{\end{equation}}
\newcommand{\ba}{\begin{eqnarray}}
\newcommand{\ea}{\end{eqnarray}}
\newcommand{\bi}{\begin{itemize}}
\newcommand{\ei}{\end{itemize}}
\newcommand{\bse}{\begin{subequations}}
\newcommand{\ese}{\end{subequations}}

\title{Probing Out-of-Time-Order Correlators}

\author[a]{Soumyadeep Chaudhuri}
\author[a]{\!, R. Loganayagam}

\vspace{2cm}

\affiliation[a]{International Centre for Theoretical Sciences (ICTS-TIFR), \\
Tata Institute of Fundamental Research, \\
Shivakote, Hesaraghatta, \\
Bangalore 560089, INDIA}

\emailAdd{soumyadeep.chaudhuri@icts.res.in}
\emailAdd{nayagam@icts.res.in}

\abstract{We present a method to probe the Out-of-Time-Order Correlators (OTOCs) of a general system by coupling it to a harmonic oscillator probe.  When the system's degrees of freedom are traced out, the OTOCs imprint themselves on the generalized influence functional of the oscillator. This generalized influence functional leads to a local effective action for the probe whose couplings encode OTOCs of the system. We study the structural features of this effective action and the constraints on the couplings from microscopic unitarity. We comment on how the OTOCs of the system appear in the OTOCs of the probe.}

\begin{document}
\maketitle

\section{Introduction}

In the study of a quantum system, a common question  of interest is how does it evolve when perturbed from an initial state. The answer lies in the future response of the system which is encoded in expectation values of strings of operators ordered in time (time-ordered correlators). However, for a variety of questions, such time-ordered correlators are no more adequate. 

For example, say we wanted to quantify the chaotic behaviour in quantum evolution. This  question is naturally addressed by imagining the following : first, we create a state of the system at a particular time instant (the present). Then we evolve the system backward in time and add a perturbation in its past. Next, we evolve it forward to the present and examine how much this procedure has modified its state. When translated into correlators, this leads us naturally to correlation functions that violate time-ordering.

Such Out-of-Time-Order Correlators (OTOCs) have received much attention \cite{Shenker:2014cwa,Maldacena:2015waa,Swingle:2016var,2016arXiv160701801Y,PhysRevLett.119.026802,Swingle:2017jlh, Menezes:2017ulq, Bohrdt:2016vhv, Knap:2018pmj,2016arXiv160801914F, 2017AnP...52900332C, 2017AnP...52900318H,2018arXiv180401545R, 2017NatPh..13..781G,PhysRevX.7.031011} recently, both from theoretical and experimental points of view. Most of the attention has been focused on a particular class of 4-point thermal OTOCs which are of the following form:
\be
\langle W(t)^\dag V(0)^\dag W(t)V(0)\rangle\ ,
\ee
where W and V are two operators acting on the Hilbert space of the system.
In chaotic systems, such thermal OTOCs show an exponentially fast fall-of in a certain time-regime \cite{Roberts:2014isa, Shenker:2014cwa, Maldacena:2015waa}. The rate of this fall-off is treated as a quantum counterpart of the Lyapunov exponent in classical chaotic systems.

Such Lyapunov exponents for thermal states in quantum systems have been shown to be constrained by an upper bound \cite{Maldacena:2015waa}.\footnote{The upper bound on the Lyapunov exponent is given by $\frac{2\pi}{\beta}$ where $\beta$ is the inverse temperature (in units where $k_B=1$ and $\hbar=1$) of the system.} Moreover, for large N gauge theories which have holographic duals, the Lyapunov exponent saturates this bound \cite{Shenker:2014cwa, Maldacena:2015waa}. This observation has led to a burst of interest in such diagnosis of chaos through OTOCs as they can provide hints to find simple quantum systems \cite{Maldacena:2016hyu, Witten:2016iux, Klebanov:2016xxf} which may have holographic duals.  Such systems can then be used as toy models to gain valuable insights into the structure of quantum theories of gravity \cite{Mandal2017, Das:2017pif, Das:2017wae, Gaikwad:2018dfc}.

Despite these progresses in the study of OTOCs, we still lack an intuitive picture of these correlations and many familiar tools of effective theory are yet to be extended to include the information contained in them. In this work, we begin to address this issue by  asking how OTOCs of a system get imprinted on the effective theory of its probe. We also develop a systematic formalism  to extend various ideas familiar in the theory of open quantum systems to OTOCs.

One main obstacle in understanding  OTOCs  is the difficulty of computing them. If the system is strongly coupled then  perturbative techniques  cannot be employed  to calculate the correlators. Even in the case of weakly coupled systems, the diagrammatic techniques \cite{Aleiner:2016eni,Stanford:2015owe,Maldacena:2016hyu} for computing OTOCs are far less developed than their counterparts for time-ordered correlators. The reason for additional complication in case of OTOCs is the proliferation of Feynman diagrams due to the multiplicity of fields, propagators and vertices. The problem becomes acute particularly when the system has many degrees of freedom with complicated interactions between them.
  
This problem of computing OTOCs in large systems becomes tractable if we restrict our attention to operators acting on a small subset of degrees of freedom. One can then take these  degrees of freedom to define an open quantum system (See \cite{PhysRevB.97.161114} and \cite{2018PhRvL.120t1604D} for particular examples) and try to write down an effective theory for its evolution. Such an effective theory can immensely simplify computations of OTOCs.

A paradigmatic example of such an open quantum system is a quantum Brownian particle interacting with a general environment. We can then think of the environment as the large system that the particle is probing. A systematic path integral formalism to understand such a system was developed by  Feynman and Vernon \cite{Feynman:1963fq} which was  then applied to a simple example of a harmonic bath.
For the (bath+particle) combined system, they integrated out the bath's degrees of freedom to get an \textit{influence phase}\footnote{The influence phase is $(-i)$ times the logarithm of the influence functional. It determines the evolution of the reduced density matrix of the particle including the physics of its decoherence.} for the particle.  This analysis was extended by Caldeira and Leggett in \cite{Caldeira:1982iu} and by Hu, Paz and Zhang in \cite{PhysRevD.45.2843, PhysRevD.47.1576} for various thermal baths and for different kinds of particle-bath couplings. These analyses can equivalently be understood in the Schwinger-Keldysh formalism \cite{Schwinger:1960qe, Keldysh:1964ud,CHOU19851,kamenev_2011}. These works however do
not explain  how OTOCs get communicated and how they  decohere within quantum systems.  

In this work, we obtain an effective theory for the probe in the generalized Schwinger-Keldysh formalism\cite{Aleiner:2016eni, 2017arXiv170102820H}. This facilitates the computation of the probe's OTOCs which are determined in terms  of the effective couplings. These couplings, as we will see, encode partial information about the OTOCs of the system. This opens up the interesting possibility that measuring the OTOCs of the probe might be a way to access the OTOCs of a large system. It would be interesting to extend the currently existing OTOC measurement protocols \cite{2017NatPh..13..781G,PhysRevX.7.031011,Zhu:2016uws,Swingle:2016var,2016arXiv160701801Y, Bordia:2018wnc, 2017PhRvL.119d0501G} to this context. The effective action described in this work might also be relevant in describing decoherence in the context of weak measurements\cite{PhysRevLett.60.1351,PhysRevA.95.012120, Halpern:2017abm, PhysRevA.98.012132,Swingle:2018xvb}.

In the effective theory of the probe worked out in this paper, we take into account the contribution of connected parts of 3-point OTOCs of the system. We choose to deal with this  case, rather than studying the effects of the 4-point OTOCs, because all the essential ideas that go into the construction of the framework can be developed in this simpler context. In section \ref{conclusion}, we comment on the possible extension of the framework to incorporate the effects of the 4-point OTOCs of the system. As mentioned in section \ref{conclusion}, such an extension can be useful to extract information about the Lyapunov exponents in chaotic systems from the parameters in the effective theory of the probe.\footnote{See \cite{2018PhRvL.120t1604D} for a concrete example where the Lyapunov exponent of a system is related to a similar exponent in the dynamics of a probe. This example involves a maximally chaotic system which has a holographic dual. It would be interesting to see whether the effective theory framework introduced here can be used to obtain similar relations for systems showing sub-maximal chaos.}

The structure of this paper is as follows : we begin with a simple example of a probe and discuss its coupling to the system. The OTOCs of this combined system are captured by a generalized Schwinger-Keldysh path integral. Integrating out the system's degrees of freedom  in this path integral results in the generalized influence phase for the probe which can be used to obtain a non-local non-unitary 1-PI  effective action.\footnote{The tree level diagrams of the 1-particle irreducible effective action provide the full perturbative expansion of the probe's correlators in the system-probe coupling.} In subsequent sections, we restrict ourselves to a local/Markovian limit. The dynamics of the probe in such a limit is described by a local 1-PI effective action whose form is constrained by the unitarity of the combined system. The couplings in this effective action are determined in terms of the OTOCs of the system.  The OTOCs of the probe are in turn determined in terms of these couplings.

\section{Specification of the probe}
Consider a quantum system S.  Let $O(t)$ be an operator (in the Heisenberg picture) acting on the Hilbert space of S. Suppose we are interested in the OTOCs of this operator. These OTOCs can be extracted from the OTOCs of a probe coupled to the system. For simplicity, we take the probe to be a harmonic oscillator of unit mass. We denote the position of the probe by $q$ and the degrees of freedom of S  collectively by $X$. The overall Lagrangian of the system and the probe is given by
\be
\begin{split}
L[q,X]=\frac{1}{2} \Big(\dot{q}^2-m_0^2q^2\Big)+L_{S}[X]+\lambda  \ O\ q.
\end{split}
\label{combined Lagrangian}
\ee
Here, $m_0$ is the frequency of the probe, $L_{S}[X]$ is the Lagrangian of the system and $\lambda$ is the strength of interaction between the system and the probe. We will take $\lambda$ to be small i.e. the probe to be weakly coupled to S. This allows us to employ perturbation theory in obtaining an effective dynamics of the probe.

For definiteness, we will assume that the system and the probe are initially unentangled and the interaction between them is switched on at a time $t_0$.  Thus, the density matrix of the system and the probe at time $t_0$ is given by
\be
\rho(t_0)=\rho_S(t_0) \otimes \rho_{probe} (t_0)\ ,
\ee
where $\rho_S(t_0)$ and $\rho_{probe} (t_0)$ are the initial density matrices of the sytem and the probe respectively. 

In the next section we will write down an action for the (system+probe) combined system in the generalized Schwinger-Keldysh formalism. In the corresponding path integrals we will integrate out the system's degrees of freedom to obtain a generalized influence phase for the probe which would allow computation of its OTOCs .
\section{Generalized influence phase for the probe}
Before introducing the action for the combined system, let us motivate the need for working in the generalized Schwinger-Keldysh formalism. Such a generalization is required for OTOCs with 3 or more insertions.
To be specific, let us focus on 3-point correlators of the operator \footnote{Here, we work in units where $\hbar=1$.}
\be
O(t)\equiv e^{iH_S(t-t_0)}O(t_0)e^{-iH_S(t-t_0)}
\ee
where $H_S$ is the Hamiltonian of the system. Our aim is to extract information about certain 3-point correlators which have 2 future-turning point insertions (i.e. insertions whose immediate neighbors lie to their pasts). For example, for $t_1>t_2>t_3>t_0$, the correlator 
\be
\langle O(t_1)O(t_3)O(t_2)\rangle\equiv Tr\Big(\rho_S(t_0)O(t_1)O(t_3)O(t_2)\Big)
\label{2OTOcorrelator}
\ee
 has 2 future turning point insertions: $O(t_1)$ and $O(t_2)$. The neighbors of both these  insertions are $\rho_S(t_0)$ and $O(t_3)$ which lie to their pasts. 
 A correlator with $k$ future-turning point insertions is called a $k$-OTO correlator \cite{2017arXiv170102820H}.  So, the correlator given in \eqref{2OTOcorrelator} is a 2-OTO correlator.
 
 We want to see the effects of such 2-OTO correlators of the system on the correlators of the probe. But such effects are not captured in the 1-OTO correlators of the probe. This is due to the fact that the 1-OTO correlators of the operator $O(t)$ completely determine the quantum master equation \cite{Lindblad:1975ef, BRE02} for the reduced density matrix of the probe, or equivalently, its influence phase in the Schwinger-Keldysh formalism (see \cite{Sieberer:2015svu, PhysRevD.45.2843, PhysRevD.47.1576} for how these two are related).  This influence phase in turn is sufficient to determine the 1-OTO correlators of the probe.  Hence, to see the effect of 2-OTO correlators of $O(t)$, one has to look at the 2-OTO correlators of the probe. 
 
\begin{figure}[h!]
\caption{A contour with 2 time-folds}
 \begin{center}
\scalebox{0.7}{\begin{tikzpicture}[scale = 1.5]
\draw[thick,color=violet,->] (-2,0 cm) -- (2,0 cm) node[midway,above] {\scriptsize{1}} ;
\draw[thick,color=violet,->] (2, 0 cm - 0.5 cm) -- (-2, 0 cm - 0.5 cm)node[midway,above] {\scriptsize{2}}  ;
\draw[thick,color=violet,->] (2, 0 cm) arc (90:-90:0.25);
\draw[thick,color=violet,->] (-2,-1.0 cm) -- (2,-1.0 cm) node[midway,above] {\scriptsize{3}} ;
\draw[thick,color=violet,->] (2, -1.0 cm - 0.5 cm) -- (-2, -1.0 cm - 0.5 cm) node[midway,above] {\scriptsize{4}} ;
\draw[thick,color=violet,->] (2, -1.0 cm) arc (90:-90:0.25);
\draw[thick,color=violet,->] (-2,-1.0 cm + 0.5 cm) arc (90:270:0.25);
\end{tikzpicture}}
 \label{fig:contour}
 \end{center}
\end{figure}
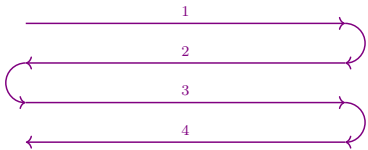

 To get a path integral representation of the 2-OTO correlators, one has to extend the Schwinger-Keldysh contour to a contour with two time folds \cite{Aleiner:2016eni, 2017arXiv170102820H,Haehl:2016pec} as shown in Figure \ref{fig:contour}.
Such a contour has four legs labelled by $1$, $2$, $3$ and $4$. We will take the past-turning point in the contour at the time $t_0$ and the future-turning points at some time T which is greater than the position of all the insertions in any correlator of our interest. 

One has to take one copy of the degrees of freedom of both the system and the probe for each leg: $\{q_1,X_1\},\{q_2,X_2\},\{q_3,X_3\}$ and $\{q_4,X_4\}$. We find it convenient to define q's on even legs with an extra minus sign over the convention followed in \cite{Haehl:2016pec, 2017arXiv170102820H}. Any operator in the combined system is a functional of $q$ and $X$. Hence, one gets four copies of such operators in this formalism.
To obtain 2-OTO correlators of these operators, one has to compute path integrals with an action given by
\be
\begin{split}
S_{\text{2-fold}}=\int_{t_0}^T dt &\Big\{L[q_{1},X_{1}]-L[-q_{2},X_{2}]+L[q_{3},X_{3}]-L[-q_{4},X_{4}]\Big\}.
\end{split}
\ee
These path integrals with insertions on any of the four legs give contour-ordered  correlators in the single-copy theory i.e. the operators in the single-copy theory corresponding to the insertions are ordered from right to left as one moves along the arrow indicated in Figure~\ref{fig:contour}.

In order to  calculate correlators of the probe, we can first integrate out the degrees of freedom of the system to  obtain a generalized influence phase \cite{Feynman:1963fq}
for the probe. This generalized influence phase $W$ can be expanded in powers of $\lambda$ as
\be
W=\lambda W_1+\lambda^2 W_2+\lambda^3 W_3+\ldots\ \ \ \ .
\label{infphase1}
\ee
For $n\geq 1$, $W_n$ is given by
\be
\begin{split}
W_n=&\frac{i^{n-1}}{n!}   \int_{t_0}^T dt_1\cdots \int_{t_0}^{T} dt_n\ \sum_{i_1,\cdots,i_n=1}^4 \langle \mathcal{T}_C O_{i_1}(t_1)\cdots O_{i_n}(t_n)\rangle_c\ q_{i_1}(t_1) \cdots q_{i_n}(t_n)\ ,
\end{split}
\label{infphase2}
\ee
where $\langle \mathcal{T}_C O_{i_1}(t_1)\cdots O_{i_n}(t_n)\rangle_c$ is the cumulant (connected part) of a contour-ordered correlator of the operators $O(t_j)$ calculated in the initial state $\rho_S(t_0)$ with the insertion at time $t_j$ on the $i_j^{\text{th}}$ leg.

With this generalized influence phase, one can calculate the OTOCs of the probe. The cumulants of such OTOCs can also be obtained from the connected tree-level diagrams of a 1-PI effective action \footnote{It is natural to wonder whether there exists an equivalent Wilsonian effective action whose  connected tree and loop-level diagrams reproduce these cumulants. The existence and the exact structure of such non-unitary Wilsonian effective actions happen to be surprisingly subtle and will be discussed elsewhere.}.This  1-PI effective action provides the Schwinger-Dyson equations satisfied by the OTOCs. These equations can then be solved to fix these OTOCs, once the necessary boundary conditions are provided. In the next section we impose some conditions on the form of this effective action.
\section{1-PI Effective Action for the probe}
\label{1-PI effective}

 The 1-PI effective action obtained from the generalized influence phase is usually non-local. But one can work in the Markovian limit \cite{BRE02} to get an approximate local form if the cumulants of correlators of $O(t)$ decay  sufficiently fast compared to the natural time-scales of the probe.  Such a local form was worked out in \cite{Caldeira:1982iu, PhysRevD.45.2843,PhysRevD.47.1576, BRE02} for the Schwinger-Keldysh effective action of a Brownian particle interacting with a variety of thermal baths. To see the regime of validity of such a local dynamics in a concrete case, let us briefly discuss a simple model which was first introduced by Caldeira-Leggett in \cite{Caldeira:1982iu} (see \cite{BRE02} for a detailed analysis of this model).

Consider  a bath made up of harmonic oscillators at inverse temperature $\beta$. Suppose the bath oscillators are interacting linearly with the position of a Brownian particle. Then the operator coupling to the particle is given by
\be
\lambda O(t)\equiv \lambda \sum_i g_i X_i(t),
\ee 
where $X_i(t)$ are the positions of the bath oscillators at time t. Let the distribution of the couplings $g_i$ be such that the  2-point spectral function of this operator has the following form:
\be
\begin{split}
\rho[12] & \equiv \lambda^2\int_{-\infty}^\infty dt_1\int_{-\infty}^\infty dt_2\ \langle[O(t_1,O(t_2)]\rangle\ e^{i(\omega_1t_1+\omega_2t_2)}\\
&=2\pi \delta(\omega_1+\omega_2)\Big[2 M_p \gamma \frac{\omega_1 \Omega^2}{\omega_1^2+\Omega^2}\Big].
\end{split}
\label{CLspectral}
\ee
Here $M_p$ is the renormalized mass of the particle and $\Omega$ is a UV cut-off frequency. As discussed in \cite{BRE02}, one can show that the parameter $\gamma$ appearing in this spectral function is the coefficient of damping induced in the particle's dynamics due to its interaction with the bath.

For the spectral function given in \eqref{CLspectral}, there exists a regime of parameters, where the 2-point functions of O(t) decay exponentially with increase in separation between the insertions \cite{BRE02}. In the high temperature limit, where $\beta\Omega<<1$, the decay rate of these 2-point functions is given by $\Omega$  \cite{BRE02}. When this decay of the bath correlators is much faster than the particle's evolution, the effect (via the bath) of any previous state of the particle on its later dynamics is heavily suppressed. 

Therefore, we see that to get an approximately local effective theory for the particle, $\Omega$ has to be much greater than the frequency-scales involved in the particle's evolution.  These frequency-scales are set by the particle's natural frequency($m_0$) and its damping coefficient($\gamma$). Hence, one can obtain a Markovian limit of the particle's dynamics when the parameters in the (particle+bath) combined system satisfy the following hierarchy:
\be
\frac{1}{\beta}>>\Omega>>\text{max}[m_0,\gamma].
\ee

In the simple model that we discussed above, the operator coupling to the particle is linear in the positions of the bath oscillators. Hence, by Wick's theorem, the cumulants of all thermal n-point functions of this operator vanish for $n>2$. Consequently, from the form of the influence phase given in \eqref{infphase2}, we can see that the effective theory of the particle is quadratic. The effective couplings in this quadratic theory receive contributions only from 2-point functions of $O(t)$. But in this paper, we want to probe the 3-point OTO correlators of the system. To see the effects of such OTOCs, one needs to couple the particle to an operator which has non-vanishing cumulants for 3-point correlators. These cumulants would then contribute to the cubic couplings in the particle's effective theory.

One way to obtain such cubic (as well as higher degree) terms in the effective theory would be to introduce some term in $O(t)$ which is nonlinear in the position of the bath oscillators. For such particle-bath interactions, the regime of validity of Markov approximation  for the quadratic terms in the effective theory was discussed in \cite{PhysRevD.47.1576}. In a similar context, the Markovian regime for the cubic terms will be discussed in detail elsewhere \cite{Chakrabarty:2018dov}. As mentioned earlier, the argument for such local dynamics of the probe rests on the decay of the system's  cumulants being much faster than the probe's evolution. This can be achieved by taking the probe's natural frequency and the system-probe coupling to be small enough.\footnote{See the comments in section \ref{conclusion} for some further discussion on the Markovian regime.}
 
From here onwards, we will restrict our discussion to the case where both the 1-OTO and 2-OTO cumulants of the operator $O(t)$ decay sufficiently fast and consequently, a local 1-PI effective action for the probe is a valid approximation. Moreover, we assume that all cubic terms in the 1-PI effective action with more than one derivative acting on q's are negligible. For the part quadratic in the q's, we keep terms with up to two derivatives to take into account the kinetic term.

The local 1-PI effective action for the probe should satisfy certain conditions 
which are based on the following two facts:
\begin{enumerate}[a)]
 \item the probe is a part of a closed system described by a unitary dynamics, 
 \item the operator q is Hermitian. 
 \end{enumerate}
 We  enumerate these conditions below:
\begin{enumerate}[ref={condition~\arabic*}]
\item \underline{Collapse Rules}

The 1-PI effective action becomes independent of $\tilde{q}$ under any of the following identifications:
\begin{enumerate}
\item \textbf{(1,2) collapse:} $q_1=-q_2=\tilde{q}$,
\item \textbf{(2,3) collapse:} $q_2=-q_3=\tilde{q}$, 
\item \textbf{(3,4) collapse:} $q_3=-q_4=\tilde{q}$ .
\end{enumerate}
\label{Collapse rules}
Under any of these collapses, the 1-PI effective action reduces to the Schwinger-Keldysh effective action in which the residual degrees of freedom play the role of the right-moving and the left-moving coordinates\cite{Haehl:2016pec}.
\item  \underline{Reality condition}

The 1-PI effective action should become the negative of itself under complex conjugation of all the couplings  and the following exchanges:
\begin{equation*}
q_1 \leftrightarrow -q_4,\ q_2 \leftrightarrow -q_3 \ .
\end{equation*}
\label{Reality condition}
\end{enumerate}

The first condition ensures that a contour-ordered correlator of the probe just picks up a sign if one slides an insertion from one leg to its adjacent leg at the same temporal position without encountering any obstruction from other insertions.\footnote{See \cite{2017arXiv170102820H} for discussions on such relations between contour-ordered correlators.} The change in sign of the correlator is due to our choice of putting an extra minus sign for q's on the left-moving legs  over the standard convention usually followed in the literature on the Schwinger-Keldysh formalism \cite{CHOU19851,2009AdPhy..58..197K, Haehl:2016pec} and its generalization to contours  with multiple time-folds \cite{Haehl:2016pec, 2017arXiv170102820H}.

 The second condition is necessary to ensure that correlators with insertions of Hermitian operators in opposite orders are complex conjugates of each other.  
 
 These conditions are straightforward extensions of the conditions imposed on the Schwinger-Keldysh effective action without any term involving  derivatives in \cite{Avinash:2017asn}. 

We will write down a local 1-PI effective Lagrangian consistent with the above conditions which has the following expansion:
\be
L_{\text{1PI}}=L_{\text{1PI}}^{(1)}+L_{\text{1PI}}^{(2)}+L_{\text{1PI}}^{(3)}+\cdots\ ,
\ee
where the $L_{\text{1PI}}^{(1)}$, $L_{\text{1PI}}^{(2)}$ and $L_{\text{1PI}}^{(3)}$ are the terms linear, quadratic and cubic in q's respectively. The linear and quadratic terms are given in \eqref{1PIeff1234:linear} and \eqref{1PIeff1234:quadratic} respectively.
\be
L_{\text{1PI}}^{(1)}=F(q_1+q_2+q_3+q_4)\ ,
\label{1PIeff1234:linear}
\ee
\be
\begin{split}
L_{\text{1PI}}^{(2)}&=\frac{1}{2} Z (\dot{q}_1^2+\dot{q}_3^2)-\frac{1}{2} Z^* (\dot{q}_2^2+\dot{q}_4^2)+i\ Z_\Delta\sum_{i<j}\dot{q}_i \dot{q}_j\\
&\quad-\frac{m^2}{2}(q_1^2+q_3^2)+\frac{(m^2)^*}{2}(q_2^2+q_4^2)\\
&\quad-i m_{\Delta}^2 \sum_{i<j}q_i q_j+\frac{\gamma}{2} \sum_{i<j} ({q_i} \dot{q}_j-\dot{q_i} {q_j})\ .
\end{split}
\label{1PIeff1234:quadratic}
\ee

The cubic terms can be split into 2 parts: One part which reduces to the terms in the Schwinger-Keldysh 1-PI effective action under any of the collapses mentioned above, and another part which vanishes under such collapses. These 2 sets of terms are given in \eqref{1PIeff1234:cubicSK} and \eqref{1PIeff1234:cubicOTO}.
\be
L_{\text{1PI}}^{(3)}=L_{\text{1PI,SK}}^{(3)}+L_{\text{1PI,2-OTO}}^{(3)}\ ,
\ee
where
\be
\begin{split}
L_{\text{1PI,SK}}^{(3)}&=-\frac{\lambda_3}{3!} (q_1^3+q_3^3)-\frac{\lambda_3^*}{3!} (q_2^3+q_4^3)\\
&\quad+\frac{\sigma_3}{2!}\Big[q_1^2(q_2+q_3+ q_4)- q_2^2 (q_3+q_4)+q_3^2  q_4\Big]\\
& \quad+\frac{\sigma_3^*}{2!}\Big[q_1(q_2^2-q_3 ^2+q_4^2)-q_2(q_3^2-q_4^2)+q_3 q_4^2\Big]\\
&\quad+\frac{\sigma_{3 \gamma}}{2!} \Big[q_1^2 (\dot{q}_2+\dot{q}_3+\dot{q}_4)-q_2^2(\dot{q}_3+\dot{q}_4)+q_3^2 \dot{q}_4-(q_2^2\dot{q}_2+q_4^2\dot{q}_4)\Big]\\
&\quad+\frac{\sigma_{3 \gamma}^*}{2!} \Big[\dot{q}_1 (q_2^2-q_3^2+q_4^2)-\dot{q}_2({q}_3^2-{q}_4^2)+\dot{q}_3 {q}_4^2-(q_1^2\dot{q}_1+q_3^2\dot{q}_3)\Big],
\end{split}
\label{1PIeff1234:cubicSK}
\ee

\be
\begin{split}
&L_{\text{1PI,2-OTO}}^{(3)}\\
&=-\Big(\kappa_3+\frac{1}{2} \text{Re}[\lambda_3-\sigma_3]\Big) (q_1+q_2)(q_2+q_3)(q_3+q_4)\\
&\quad-({q}_2+{q}_3) \Big[\Big(\kappa_{3\gamma}- \text{Re}[\sigma_{3 \gamma}]\Big) (\dot{q}_1+\dot{q}_2) ({q}_3+{q}_4)\\
&\qquad\qquad\qquad+\Big(\kappa_{3\gamma}^*- \text{Re}[\sigma_{3 \gamma}]\Big) ({q}_1+{q}_2) (\dot{q}_3+\dot{q}_4)\Big]\ .
\end{split}
\label{1PIeff1234:cubicOTO}
\ee
The collapse rules further impose the following conditions \cite{Avinash:2017asn} on the couplings:
\be
\begin{split}
& Z_\Delta=\text{Im}[Z],\ m_{\Delta}^2 = \text{Im}[m^2],\ \text{Im}[\lambda_3+3 \sigma_3]=0\ .
\end{split}
\label{Lindblad}
\ee
The reality condition implies that $F,\ \gamma$ and $\kappa_3$ are real.

The terms given in \eqref{1PIeff1234:linear}, \eqref{1PIeff1234:quadratic} and  \eqref{1PIeff1234:cubicSK} are extensions of the terms appearing in the Schwinger-Keldysh(SK) effective theory (see appendix \ref{app:SKeffaction} for the explicit form of the SK effective action). This is the most general local SK effective action up to cubic terms with a single derivative. The cubic terms without any derivative were discussed in the context of an open scalar field theory in \cite{Avinash:2017asn}. To our knowledge, this is the first work which identifies all the possible cubic terms with a single derivative in the SK effective theory. It also introduces, for the first time, an extension of such terms in the SK effective theory to the effective theory on the 2-fold contour.

The quadratic terms in the SK effective theory  have been studied previously in detail by Caldeira-Leggett \cite{Caldeira:1982iu} and Hu-Paz-Zhang \cite{PhysRevD.45.2843, PhysRevD.47.1576}. This quadratic effective theory has been shown to be equivalent to a stochastic dynamics governed by a linear Langevin equation with a Gaussian noise \cite{kamenev_2011} following the methods developed by Martin-Siggia-Rose \cite{PhysRevA.8.423}, De Dominicis-Peliti \cite{PhysRevB.18.353} and Janssen \cite{Janssen1976}. In this Langevin dynamics, the coupling $\gamma$ is the coefficient of damping, whereas $m_\Delta^2$ is the strength of the noise experienced by the particle. The real parts of $m^2$ and $Z$ are the renormalized frequency (squared) and mass of the particle respectively. The linear term in \eqref{1PIeff1234:linear} augments this dynamics by introducing a constant force on the particle. \footnote{The correction to the Langevin dynamics induced by the terms  corresponding to $Z_\Delta=\text{Im}[Z]$ will be discussed elsewhere \cite{Chakrabarty:2018dov}.}  

Among the cubic terms given in \eqref{1PIeff1234:cubicSK}, the term associated with the real part of $\lambda_3$ is the usual cubic potential in a unitary dynamics of the particle.  The non-unitary couplings $\sigma_3$ and $\sigma_{3\gamma}$ are coefficients of terms that mix the degrees of freedom on different legs. These cubic terms introduce a nonlinearity in the equivalent stochastic theory. A detailed treatment of this nonlinear Langevin dynamics will be done elsewhere \cite{Chakrabarty:2018dov}.

The couplings $\kappa_3$ and $\kappa_{3\gamma}$ are not present in the Schwinger-Keldysh 1-PI effective action. They encode information about the 2-OTO 3-point functions of the operator $O(t)$ (see Table \ref{coupling-correlator}). Hence, these couplings are of central importance in this paper. As we will see, to determine these couplings one needs to measure some 2-OTO correlators of the probe (see equations \eqref{nestedcommprobe} and \eqref{2OTOdiff}). 
\section{Relations between 1-PI effective couplings and system's correlators}
 The couplings in the 1-PI effective action can be derived from the generalized influence phase given in \eqref{infphase2} (see appendix \ref{app:coup-corr} for an outline of the arguments involved in the derivation). These couplings will generally be functions of time. But we focus on the particle's dynamics at a sufficiently late time when the effective couplings have saturated to constant values. 
 Moreover, we assume that
\be
 \lim_{t-t_0\rightarrow\infty}\langle O(t)\rangle=0.
 \label{late time 1-pt O}
\ee
If this is not true, then one can give a constant shift to the centre of oscillation of the probe. This effectively introduces a shift in the operator $O(t)$ when the Lagrangian is recast into the form given in \eqref{combined Lagrangian}. By appropriately choosing this shift, one can make sure that the condition given in  \eqref{late time 1-pt O} is satisfied by the shifted operator. This condition implies that the O$(\lambda)$ term in the linear coupling vanishes. However, we expect a subleading contribution at O($\lambda^3$). A correct computation of this subleading term requires taking into account the perturbative corrections to the state of the probe. We will consistently ignore such subleading corrections in what follows.
 
{We restrict our attention to the relations connecting the leading order forms of the quadratic and the cubic couplings to the correlators of the operator $O(t)$. These relations are given in equations \eqref{quadratic coupling}, \eqref{cubic-linear} and Table \ref{coupling-correlator}. 

\paragraph{Notational conventions:}
While expressing the couplings in terms of the correlators of $O(t)$, we have followed some notational conventions which are given below:
\begin{enumerate}
\item The interval between two time instants $t_i$ and $t_j$ is expressed as
\be
t_{ij}\equiv t_i-t_j.
\ee
\item We express the cumulant 
$\langle O(t_{i_1})O(t_{i_2}) \cdots O(t_{i_n}) \rangle_c$ as $\langle i_1 i_2 \cdots i_n \rangle$. For example,
\be
\langle 123 \rangle\equiv\langle O(t_{1})O(t_{2})O(t_{3}) \rangle_c\ .
\ee
\item The cumulant corresponding to a single-nested structure with commutators and anti-commutators is expressed by angle brackets enclosing a pair of square brackets\cite{Haehl:2017eob}. The insertions that one encounters while going outwards through the nested structure are arranged from left to right within the square brackets.
Positions of anti-commutators are indicated by (+) signs. For example,
\be
\begin{split}
\langle [123] \rangle&\equiv\langle [[O(t_{1}),O(t_{2})],O(t_{3})] \rangle_c\ ,\\
\langle [12_+3] \rangle&\equiv\langle [\{O(t_{1}),O(t_{2})\},O(t_{3})] \rangle_c\ ,\\
\langle [321_+] \rangle&\equiv\langle \{[O(t_{3}),O(t_{2})],O(t_{1})\} \rangle_c\ .
\end{split}
\ee
\end{enumerate}

\paragraph{Quadratic couplings:}
\be
\begin{split}
Z&=1-i\ \lambda^2\ \lim_{t_{10}\rightarrow\infty}\Big[\int_{t_0}^{t_1}dt_2\langle 12\rangle t_{12}^2\Big]+\text{O}(\lambda^4),\\
m^2 &= m_0^2-2i\ \lambda^2\ \lim_{t_{10}\rightarrow\infty}\Big[\int_{t_0}^{t_1}dt_2\langle 12\rangle\Big]+\text{O}(\lambda^4), \\ 
\gamma&=i\ \lambda^2\ \lim_{t_{10}\rightarrow\infty}\Big[\int_{t_0}^{t_1}dt_2\langle [12]\rangle t_{12}\Big]+\text{O}(\lambda^4) .\\ 
\end{split}
\label{quadratic coupling}
\ee
\paragraph{Cubic couplings:}
Any cubic coupling $g$ can be expanded in powers of $\lambda$ as 
\be
\begin{split}
g=\lambda^3\lim_{t_{10}\rightarrow\infty}\int_{t_0}^{t_1}dt_2 \int_{t_0}^{t_2} dt_3\ \mathcal{I}[g]+\text{O}(\lambda^5)\ .
\end{split}
\label{cubic-linear}
\ee
We enumerate the integrand $\mathcal{I}$ for the cubic couplings in Table \ref{coupling-correlator}. 

\begin{table}[ht]
\caption{Relations between the probe's 1-PI effective couplings and the correlators of $O(t)$}
\label{coupling-correlator}
\begin{center}
\begin{tabular}{ |c| c|  }
\hline
$g$ & $\mathcal{I}[g]$\\
\hline
$\lambda_3$ & $6\langle 123\rangle$ \\
\hline
$\text{Re}[\lambda_3+\sigma_3]$ & $2\langle [123]\rangle$\\
\hline
$\kappa_3$ & $-\langle [321]\rangle$\\
\hline
2 \text{Re}[$\sigma_{3\gamma}$]&$- \langle [123]\rangle (t_{12}+t_{13})$\\
\hline
2 \text{Re}[$ \kappa_{3\gamma}$] & $ \langle [321]\rangle\ (t_{32}+t_{31})$\\
\hline
  2i \text{Im}[$\kappa_{3\gamma}$]&$-\Big(\langle[123_+]\rangle+\langle [321_+]\rangle\Big)t_{12}-\langle[12_+3]\rangle\ t_{13}$\\
  \hline
   2i \text{Im}[$\sigma_{3\gamma}$]&$\langle [12_+3]\rangle (t_{32}+t_{31})+ \langle[123_+]\rangle(t_{21}+t_{23})$\\
 \hline  
\end{tabular}
\end{center}
\end{table}
Notice that $\kappa_3$ and $\kappa_{3\gamma}$ are the only two cubic couplings that receive contributions from the 2-OTO cumulants $\langle 132\rangle$ and $\langle 231\rangle$ which appear in the expansions of the nested structures $\langle[321]\rangle$ and $\langle[321_+]\rangle$. In the next section, we give examples of two OTOCs of the probe where these two couplings show up.

\section{OTOCs of the probe}

The 1-PI effective action introduced in section \ref{1-PI effective} can be used to express the probe's OTOCs in terms of the effective couplings. The tree level diagrams in the effective theory provide these expressions of the OTOCs. To fix the values of the propagators in such diagrams, one needs to find the appropriate initial condition of the probe at some time when the local effective dynamics has set in. We choose this initial condition to be that of the ground state of the unperturbed oscillator (with frequency $m_0$). Given this initial state, we express the cumulants of two OTOCs of the probe in \eqref{nestedcommprobe} and \eqref{2OTOdiff}. The OTO couplings $\kappa_3$ and $\kappa_{3\gamma}$ appear in these cumulants.

While computing these OTOCs, we take the O$(\lambda^3)$ terms in the effective couplings and neglect the terms which are higher order in $\lambda$. Similarly, we take the propagators to be those corresponding to the ground state of the unperturbed oscillator and neglect  O$(\lambda^2)$ corrections to them.  This gives us the correct 3-point cumulants upto  O$(\lambda^3)$.

We find it convenient to express the time-dependence of the cumulants in terms of the phases  defined below:

\be
\phi_n\equiv \phi_0+n\Delta,\\
\ee
where
\be
\phi_0\equiv  m_0 (t_1+t_2-2t_3) ,\ \Delta \equiv m_0 (t_{3}-t_{2})\ .
\ee
For $t_1>t_2>t_3\gg t_0$, we get the following forms for the cumulants:
\be
\begin{split}
&\langle[[q(t_3),q(t_2)],q(t_1)]\rangle_c\\
&=\frac{\kappa_3}{3m_0^4}\Big\{-\cos  \phi_0+3\cos  \phi_1-3\cos  \phi_2+\cos  \phi_3\Big\}\\
&\quad+\frac{\text{Re}[\kappa_{3\gamma}]}{3m_0^3}\Big\{-2\sin \phi_0+3\sin  \phi_1-\sin\  \phi_3\Big\}+\text{O}(\lambda^5)\ ,
\end{split}
\label{nestedcommprobe}
\ee

\be
\begin{split}
&\langle q(t_1)q(t_3)q(t_2)\rangle_c-\langle q(t_2)q(t_3)q(t_1)\rangle_c\\
&=-\frac{i\ \text{Im}[2\kappa_{3\gamma}-\sigma_{3\gamma}]}{2m_0^3}\Big\{-\sin \phi_1+2\sin \phi_2-\sin \phi_3\Big\}\\
&\quad+\frac{i\ \text{Re}[\lambda_{3}+\sigma_{3}]}{6m_0^4}\Big\{-\sin (\phi_1+\phi_2)+3\sin \phi_2-\sin \phi_3\Big\}\\
&\quad+\frac{i\ \text{Re}[\sigma_{3\gamma}]}{6m_0^3}\Big\{-4\cos (\phi_1+\phi_2)+3\cos \phi_1+\cos\phi_3\Big\}+\text{O}(\lambda^5).
\end{split}
\label{2OTOdiff}
\ee
The couplings that appear in these cumulants are truncated to their leading order values in  $\lambda$ whose forms were given in \eqref{cubic-linear} and Table \ref{coupling-correlator}.

The above expressions together with Table \ref{coupling-correlator} demonstrate how the OTOCs of the probe encode information about the OTOCs of the operator $O(t)$. 
\section{Conclusion and Discussion}
\label{conclusion}
In this paper, we have demonstrated how information about the OTOCs of a generic quantum system is encoded in the OTOCs of a probe. This is done by deriving an effective action for the probe in the Markovian limit .  The couplings appearing in this effective action have been expressed in terms of the system's correlators integrated over a certain time domain. Focusing on the cubic terms in the action, we have identified the couplings that encode information about the 3-point OTOCs of the system.

We would like to emphasize that this information about the system's  OTOCs is only partial. As evident from  equation \eqref{cubic-linear} and table \ref{coupling-correlator}, the probe's effective couplings depend only on certain moments of the system's OTOCs. More information about these OTOCs can be extracted by including higher derivative terms in the probe's effective action. To determine the complete expressions for the system's OTOCs, one would have to work with the full non-local effective action of the probe.\footnote{We  thank the referee for bringing this point to our notice.}

Although we have restricted our analysis here to cubic terms in the effective action, the formalism can be extended to take into account quartic terms as well \cite{BCSC}. The corresponding couplings at leading order in $\lambda$ would receive contributions from the 4-point OTOCs of the system. These correlators have been the subject of most studies on OTOCs. It will be interesting to compare the results of such studies with those obtained from an extension of the effective theory paradigm introduced here.

While discussing the Markovian limit for the effective theory, we demanded a sufficiently fast decay of the cumulants (including the OTOCs). Such a decay can happen in several ways. For instance, the 3-point cumulants may be exponentially damped as $(e^{-\alpha_1|t_i-t_j|-\alpha_2|t_j-t_k|})$ where $t_i, t_j, t_k$ are some permutations of the time instants $t_1, t_2$, $t_3$ and $\alpha_1, \alpha_2$ are positive numbers. Such a damping of the cumulants will be discussed for a toy model elsewhere \cite{Chakrabarty:2018dov}. But for more generic systems, this kind of damping of OTOCs may not hold in all time-regimes. 

In fact, for several chaotic systems \cite{Sonner2017,PhysRevE.98.062218, PhysRevLett.121.210601,PhysRevLett.121.124101}, the 4-point OTO cumulants show an exponentially rapid fall-off (the Lyapunov regime) before saturating to some constant values. Now, if these values at which the cumulants saturate are sufficiently small and the saturation time-scales are much shorter than the time-scales at which the probe's correlators evolve, then we expect a local dynamics of the probe. The effective couplings in this dynamics would then receive contributions from the cumulants in all the different time-regimes mentioned above. It may be useful to take a simple model of a chaotic system and determine the relative significance of contributions from the different regimes to see whether information about the Lyapunov exponent can be extracted from the OTO effective couplings of the probe. 

We would like to draw attention to the fact that although we assume a weak system-probe coupling, no restriction has been imposed on the strength of couplings within the system. In particular, the system may be strongly coupled (as long as its cumulants decay sufficiently fast).  For such a system, it is not possible to employ the standard methods of perturbation theory to compute the probe's correlators directly from the microscopic dynamics. However, using the effective theory framework introduced here, one can derive the probe's correlators in terms of the effective couplings. As we have shown, these couplings encode information about the correlators of the system. Hence, a measurement of the effective couplings can provide valuable insight into the correlators of strongly coupled systems.

The effective theory formalism introduced in this paper can be extended to a large class of open quantum systems such as those relevant in the study of cavity opto-mechanics \cite{kippenberg2007cavity} or quantum optics \cite{klyshko1988photons,PhysRevA.45.6816,  PhysRevA.52.4214,PhysRevA.45.5056,Liu2018}. It can also be extended to the study of OTOCs in open quantum field theories \cite{Avinash:2017asn, PhysRevD.33.444, PhysRevD.35.495,book, PhysRevD.72.043514, Boyanovsky:2015tba,Boyanovsky:2015jen,PhysRevD.98.023515,Boyanovsky:2018fxl} which are of relevance in quantum cosmology and heavy ion physics. In this context, it will be useful to develop a Wilsonian effective theory which would allow one to study the RG flows of the OTO effective couplings and compare them with similar studies on the flows of Schwinger-Keldysh effective couplings \cite{Avinash:2017asn}.

The effective OTO dynamics of the probe  presented here holds for any generic state of the system where its cumulants decay sufficiently fast. However, it would be interesting to specialise to the case where the system is in a thermal state. 
The Kubo-Martin-Schwinger relations \cite{Haehl:2017eob} between the thermal correlators of the system would then imply additional relations between the couplings in the effective theory of the probe which are analogous to the fluctuation dissipation relations \cite{Haehl:2017eob, PhysRevE.97.012101}.  Such an effective action might also turn out to be useful  in studying the time-scale of thermalisation of  the probe's OTOCs vis a vis its time-ordered correlators. We would like to address some of these issues in future \cite{Chakrabarty:2018dov}.
\acknowledgments
RL and SC would like to thank Subhro Bhattacharjee, Bidisha Chakrabarty, Sudip Ghosh,  Gautam Mandal,  Shiraz Minwalla,  Archak Purakayastha, Mukund Rangamani, Alexandre Serantes and Douglas Stanford for useful discussions. SC is grateful to the organisers of Strings, 2018 for giving him the opportunity to present a poster on a preliminary version of this work.  He would also like to thank the participants of a discussion meeting on OTOCs at ICTS-TIFR, Bangalore for their questions and comments on this work. RL and SC are grateful for the support from International Centre for Theoretical Sciences (ICTS), Tata institute of fundamental research, Bangalore. They would also like to acknowledge their debt to the people of India for their steady and generous support to research in the basic sciences.

\appendix

\section{The Schwinger-Keldysh 1-PI effective action}
\label{app:SKeffaction}
In this appendix, we write down the form of the Schwinger-Keldysh 1-PI effective action obtained by collapsing the degrees of freedom on any two successive legs in the 1-PI effective action on the 2-fold contour.

For specificity, let us set $q_3=-q_4=\tilde{q}$ in \eqref{1PIeff1234:linear}, \eqref{1PIeff1234:quadratic} and  \eqref{1PIeff1234:cubicSK} , and impose the conditions given in \eqref{Lindblad}. Then we obtain the following SK effective action:
\be
\begin{split}
L_{SK}=& F(q_1+q_2)+\frac{\text{Re}[Z]}{2}(\dot{q}_1^2-\dot{q}_2^2)+\frac{i\ \text{Im}[Z]}{2}(\dot{q}_1+\dot{q}_2)^2-\frac{\text{Re}[m^2]}{2}(q_1^2-q_2^2)\\
&-\frac{i\ \text{Im}[m^2]}{2}(q_1+q_2)^2+\frac{\gamma}{2}(q_1\dot{q}_2-q_2\dot{q}_1)\\
&-\frac{\text{Re}[\lambda_3-3\sigma_3]}{4!}(q_1+q_2)^3-\frac{\text{Re}[\lambda_3+\sigma_3]}{8}(q_1+q_2)(q_1-q_2)^2\\
&+\frac{i\ \text{Im}[\sigma_3]}{2}(q_1+q_2)^2(q_1-q_2)-(q_1^2-q_2^2)\Big(\frac{\text{Re}[\sigma_{3\gamma}]}{2}(\dot{q}_1-\dot{q}_2)-\frac{i\ \text{Im}[\sigma_{3\gamma}]}{2}(\dot{q}_1+\dot{q}_2)\Big).
\end{split}
\ee 
Here, $q_1$ and $q_2$ are the degrees of freedom on the right and the left moving legs respectively. Notice that this effective action is independent of $\tilde{q}$ as required by the collapse rules.

One can take a linear combination of these degrees of freedom to go to the Keldysh basis:
\be
q_a\equiv \frac{1}{2}(q_1-q_2),\ q_d\equiv (q_1+q_2). 
\ee 
In this basis, the SK effective action has the following form:
\be
\begin{split}
L_{SK}=& F q_d+\text{Re}[Z]\dot{q}_a\dot{q}_d+\frac{i\ \text{Im}[Z]}{2}\dot{q}_d^2-\text{Re}[m^2]q_a q_d-\frac{i\ \text{Im}[m^2]}{2}q_d^2+\frac{\gamma}{2}(q_a\dot{q}_d-q_d\dot{q}_a)\\
&-\frac{\text{Re}[\lambda_3-3\sigma_3]}{4!}q_d^3-\frac{\text{Re}[\lambda_3+\sigma_3]}{2}q_d q_a^2+i\ \text{Im}[\sigma_3]q_d^2 q_a\\
&-2\text{Re}[\sigma_{3\gamma}] q_d q_a\dot{q}_a+i\ \text{Im}[\sigma_{3\gamma}]q_a q_d\dot{q}_d.
\end{split}
\ee 
Notice that the above action is consistent with the Lindblad condition \cite{Avinash:2017asn} i.e. there is no term which is independent of the difference field $q_d$. This is necessary to ensure the vanishing of any correlator where the difference field is the future-most insertion $-$ a condition which is based on the unitarity of the underlying microscopic theory of the system and the probe. 

\section{1-PI effective couplings from the generalized influence phase}
\label{app:coup-corr}
In this appendix, we give an outline of the arguments for the relations (see \eqref{quadratic coupling}, \eqref{cubic-linear} and table \ref{coupling-correlator}) between the effective couplings and the correlators of the operator $O(t)$. As we have already mentioned, the probe correlators can be computed from the generalized influence phase given in \eqref{infphase1} and \eqref{infphase2}.
The cubic terms in this influence phase are as follows
\be
\begin{split}
\lambda^3 W_3&=\frac{i^2\lambda^3}{3!}\int_{t_0}^T dt_1 \int_{t_0}^T dt_2 \int_{t_0}^T dt_3 \sum_{i,j ,k=1}^4 \langle \mathcal{T}_C O_{i}(t_1)O_{j}(t_2) O_{k}(t_3)\rangle_c\ q_{i}(t_1) q_{j}(t_2) q_{k}(t_3).\\
\end{split}
\ee

In the above integral, the contributions of the different domains corresponding to different orderings of the time instants $t_1,t_2$ and $t_3$ are all the same. This can be seen by simultaneously relabelling the time instants and the leg indices to go from one domain to another. So, we may restrict the integral to one such domain(say, $t_1\geq t_2\geq t_3$) and multiply by the total number of domains cancelling the $3!$ in the denominator:
\be
\begin{split}
\lambda^3 W_3 &=-\lambda^3\int_{t_0}^T dt_1 \int_{t_0}^{t_1} dt_2 \int_{t_0}^{t_2} dt_3\sum_{i,j,k=1}^4 \langle \mathcal{T}_C O_{i}(t_1)O_{j}(t_2) O_{k}(t_3)\rangle_c\ q_{i}(t_1) q_{j}(t_2) q_{k}(t_3).\\
\end{split}
\ee
Let us define the coefficient functions that multiply with the q's in the above expression as 
\be
C_{ijk}(t_1,t_2,t_3)\equiv \langle\mathcal{T}_C O_{i}(t_1)O_{j}(t_2) O_{k}(t_3)\rangle_c\ .
\ee
Then 
\be
\begin{split}
\lambda^3 W_3&=-\lambda^3\int_{t_0}^T dt_1 \int_{t_0}^{t_1} dt_2 \int_{t_0}^{t_2} dt_3\sum_{i,j,k=1}^4 C_{ijk}(t_1,t_2,t_3)\ q_{i}(t_1) q_{j}(t_2) q_{k}(t_3).\\
\end{split}
\ee

We have assumed that these coefficient functions decay much faster than the time-scales in which the correlators of q change significantly. So, we can approximate the path integral by Taylor expanding $q_{j}(t_2)$ about $(t_1-\varepsilon)$ and $q_{k}(t_3)$ about $(t_1-2\varepsilon)$ where $\varepsilon$ is a small positive number which serves as a point-split regulator in the subsequent computations. Retaining only the terms with at most a single time derivative, we get

\be
\begin{split}
\lambda^3 W_3 =-\lambda^3&\sum_{i,j ,k=1}^4\Big[ \int_{t_0}^T dt_1\Big\{ \int_{t_0}^{t_1} dt_2 \int_{t_0}^{t_2} dt_3 \ C_{ijk}(t_1,t_2,t_3)\Big\}\ q_{i}(t_1) q_{j}(t_1-\varepsilon) q_{k}(t_1-2\varepsilon)\\
&+\int_{t_0}^T dt_1 \Big\{\int_{t_0}^{t_1} dt_2 \int_{t_0}^{t_2} dt_3 \ C_{ijk}(t_1,t_2,t_3)\ t_{21}\Big\}\ q_{i}(t_1) \dot{q}_{j}(t_1-\varepsilon) q_{k}(t_1-2\varepsilon)\\
&+\int_{t_0}^T dt_1\Big\{ \int_{t_0}^{t_1} dt_2 \int_{t_0}^{t_2} dt_3\  C_{ijk}(t_1,t_2,t_3)\ t_{31}\Big\}\ q_{i}(t_1)  q_{j}(t_1-\varepsilon)\dot{q}_{k}(t_1-2\varepsilon)\Big].\\
\end{split}
\label{local_cubic_inf_phase}
\ee

These cubic vertices in the generalized influence phase contribute to connected parts of 3-point correlators of $q$ at leading order in $\lambda$. On the other hand, the same leading order forms can be obtained from the 1-PI effective action given in \eqref{1PIeff1234:cubicSK} and \eqref{1PIeff1234:cubicOTO} by truncating the expansion of the cubic couplings in $\lambda$ at the leading order. Comparing these two forms, one can express the cubic couplings in terms of the cumulants of the operator O(t) as given in \eqref{cubic-linear} and table \ref{coupling-correlator}. While making these comparisons, we ignore the contributions of terms in the generalized influence phase which are total derivatives as they can be absorbed in the boundary conditions one needs to impose on the path integral. Note that such subtractions of total derivatives would not change the equations of motion (the Schwinger-Dyson equations) of the correlators.

The expressions for the quadratic couplings given in \eqref{quadratic coupling} can be derived in a similar manner from the quadratic terms in the influence phase.



%
%

\bibliographystyle{JHEP}
\bibliography{ProbingOTOCbblg}

\end{document}